\documentclass[aps,prl,reprint,groupedaddress,showpacs,amsfonts,amsmath,amssymb,superscriptaddress]{revtex4-1}
\usepackage{graphicx}
\usepackage{bm}
\usepackage{color}

\begin{document}

\title{Spectral properties of the square-lattice antiferromagnetic $J_{1}$-$J_{2}$ Heisenberg model: confinement and deconfinement of spinons}

\author{Shun-Li Yu}
\email{slyu@nju.edu.cn}
\affiliation{National Laboratory of Solid State Microstructures and Department of Physics, Nanjing University, Nanjing 210093, China}
\affiliation{Collaborative Innovation Center of Advanced Microstructures, Nanjing University, Nanjing 210093, China}
\author{Wei Wang}
\affiliation{National Laboratory of Solid State Microstructures and Department of Physics, Nanjing University, Nanjing 210093, China}
\author{Zhao-Yang Dong}
\affiliation{National Laboratory of Solid State Microstructures and Department of Physics, Nanjing University, Nanjing 210093, China}
\author{Zi-Jian Yao}
\affiliation{Department of Physics, Nanjing Normal University, Nanjing 210023, China}
\author{Jian-Xin Li}
\email{jxli@nju.edu.cn}
\affiliation{National Laboratory of Solid State Microstructures and Department of Physics, Nanjing University, Nanjing 210093, China}
\affiliation{Collaborative Innovation Center of Advanced Microstructures, Nanjing University, Nanjing 210093, China}

\date{\today}

\begin{abstract}
Based on the mapping between $s=1/2$ spin operators and hard-core bosons, we extend the cluster perturbation theory to spin systems and study the whole excitation spectrum of the antiferromagnetic $J_{1}$-$J_{2}$ Heisenberg model on the square lattice. In the N\'{e}el phase for $J_{2}\lesssim0.4J_{1}$, in addition to the dominant magnon excitation, there is an obvious continuum close to $(\pi,0)$ in the Brillouin zone indicating
the deconfined spin-1/2 spinon excitations. In the stripe phase for $J_{2}\gtrsim0.6J_{1}$, we find similar high-energy two-spinon continuums at $(\pi/2,\pi/2)$ and $(\pi/2,\pi)$, respectively. The intermediate phase is characterized by a spectrum with completely deconfined broad continuum, which is attributed to a $Z_{2}$ quantum spin liquid with the aid of a variational-Monte-Carlo analysis.
\end{abstract}

\maketitle

The spin-$1/2$ antiferromagnetic (AF) $J_{1}$-$J_{2}$ Heisenberg model on two-dimensional square lattice has generated enormous interest due to its close relation to the magnetism in high-$T_{c}$ superconducting materials\cite{science.235.4793.1196,RevModPhys.78.17,NPHYS2438} and its possibility of realizing the so-called quantum spin liquid state. The variables $J_{1}$ and $J_{2}$ denote the nearest-neighbor (NN) and next-nearest-neighbor (NNN) exchange interactions, respectively. In the region for small $J_{2}$ ($\lesssim0.4J_{1}$), the ground state is widely believed to have a $(\pi,\pi)$ N\'{e}el AF long-range order, while the $(0,\pi)$ and $(\pi,0)$ stripe AF orders are stabilized for large $J_{2}$ ($\gtrsim0.6J_{1}$). In the intermediate region ($0.4J_{1}\lesssim J_{2}\lesssim0.6J_{1}$), many numerical studies have shown that it is a magnetically disordered and spin-rotation-invariant quantum phase. While, there is controversy on the nature of this nonmagnetic phase, and various candidate states have been proposed, such as the quantum spin-liquid (QSL)states\cite{PhysRevLett.87.097201,PhysRevLett.91.067201,PhysRevB.86.024424,PhysRevB.86.075111,PhysRevLett.111.037202,PhysRevB.88.060402}, the plaquette valence-bond state\cite{PhysRevB.54.9007,PhysRevLett.84.3173,PhysRevLett.91.197202,PhysRevB.74.144422,PhysRevB.79.024409,PhysRevB.85.094407,PhysRevB.89.104415,PhysRevLett.113.027201}, or the columnar valence-bond state\cite{PhysRevB.41.9323,PhysRevB.44.12050,PhysRevB.60.7278,PhysRevB.73.184420}. Also, the previous studies mostly focus on the ground state, while the results about the magnetic excitation spectrum are still lacking. The dynamic spectra can provide important information to identify the nature of the ground state, and it is directly related to the spectroscopic experiments such as inelastic neutron scatterings (INS).

For a magnetically ordered ground state, it is believed that the low-energy excitation spectra is well captured by
the linear spin-wave theory, in which the spin is regarded as a classical three-component vector and the excitations are quantized quasiparticles (magnon with spin-$1$) of waves due to the spin rotations.
However, a recent INS experiment on the N\'{e}el antiferromagnet $\mathrm{Cu(DCOO)}_{2}\cdot\mathrm{4D}_{2}\mathrm{O}$ (CFTD), which is considered as the best realization of the square-lattice Heisenberg model to date\cite{PhysRevLett.82.3152,PhysRevLett.87.037202,pnas.0703293104},
observes a non-spin-wave continuum at $(\pi,0)$ in the Brillouin zone (BZ)\cite{nphys3172}. This has been interpreted as a sign of deconfinement of spinons, i.e., the magnon fractionalizes into two independent $s=1/2$ objects\cite{nphys3172}. In contrast, the $(\pi/2,\pi/2)$ excitation remains more magnonlike. A recent theoretical study based on quantum Monte Carlo simulations supports this deconfinement scenario\cite{PhysRevX.7.041072}, but some other works ascribe the continuum at $(\pi,0)$ to multimagnon processes\cite{PhysRevB.52.15695,Phys.Rev.Lett.86.528,SciPost.Phys.4.001}. Thus, the nature of the continuum at $(\pi,0)$ is still under debate.

In this paper, we extend the cluster perturbation theory (CPT) to spin systems by using the mapping between $s=1/2$ spin operators and hard-core bosons. With this method, we calculate the whole dynamic spectrum of the $J_{1}$-$J_{2}$ model and clearly show the confinement and deconfinement of spinons in various phases with the variation of $J_{2}$. In the N\'{e}el phase ($J_{2}\lesssim0.4J_{1}$), in addition to the dominant magnon excitaion, we obtain an obvious continuum coming from two-spinon excitations at high energy close to $(\pi,0)$, which is consistent with the recent experimental observation\cite{nphys3172}. In the stripe phase ($J_{2}\gtrsim0.6J_{1}$), similar high-energy continuums are also found, but their locations move to $(\pi/2,\pi/2)$ and $(\pi/2,\pi)$. In the intermediate phase ($0.4J_{1}\lesssim J_{2}\lesssim0.6J_{1}$), the whole spectrum becomes a broad continuum, and we find that its characteristics are in good agreement with a $Z_{2}$ QSL based on a variational-Monte-Carlo (VMC) analysis.

As mentioned above, we consider the square-lattice spin-$1/2$ AF $J_{1}$-$J_{2}$ Heisenberg model:
\begin{align}
\mathcal{H}=J_{1}\sum_{\langle ij\rangle}\bm{S}_{i}\cdot\bm{S}_{j}+J_{2}\sum_{\langle\langle ij\rangle\rangle}\bm{S}_{i}\cdot\bm{S}_{j},
\label{eq1}
\end{align}
where $\langle ij\rangle$ and $\langle\langle ij\rangle\rangle$ denote the NN and NNN bonds, respectively. To investigate strongly correlated systems, one of the most reliable numerical methods is the exact diagonalization (ED). However, the system size that the ED can handle is too small to get the dispersion of excitations. To amend this shortcoming, we use the CPT, in which the short-range correlations are captured exactly by the ED of a small cluster and the properties of the infinite lattice are obtained with a perturbative trement of the inter-cluster couplings. This method has been successfully applied to various correlated electronic systems\cite{Phys.Rev.Lett.84.522,PhysRevLett.85.2585,PhysRevLett.92.126401,PhysRevLett.107.010401,PhysRevB.84.064520}. Here, we extend the CPT to the $s=1/2$ spin models.

Firstly, using the exact mapping between $s=1/2$ spin operators and hard-core bosonic operators\cite{ProgTheorPhys.16.569,Sov.Phys.JETP.60.781},
\begin{align}
S^{+}_{i}=b^{\dag}_{i},\quad S^{-}_{i}=b_{i},\quad S^{z}_{i}=b^{\dag}_{i}b_{i}-\frac{1}{2},
\label{mapping}
\end{align}
where $b^{\dag}$ and $b$ are the creation and annihilation operators of the hard-core boson, we rewrite the spin Hamiltonian (\ref{eq1}) as
\begin{align}
H&=\frac{1}{2}J_{1}\sum_{\langle ij\rangle}(b_{i}^{\dag}b_{j}+h.c.)+\frac{1}{2}J_{2}\sum_{\langle\langle ij\rangle\rangle}(b_{i}^{\dag}b_{j}+h.c.) \nonumber \\
&+J_{1}\sum_{\langle ij\rangle}n_{i}n_{j}+J_{2}\sum_{\langle\langle ij\rangle\rangle}n_{i}n_{j}-2(J_{1}+J_{2})\sum_{i}n_{i}.
\label{eq2}
\end{align}
The hard-core bosonic operators satisfy the commutation
relations
\begin{align}
[b_{i},b_{j}]=[b_{i}^{\dag},b_{j}^{\dag}]=0,\quad [b_{i},b_{j}^{\dag}]=\delta_{ij}(1-2b_{i}^{\dag}b_{i}),
\end{align}
and the hard-core constraint $n_{i}=0$ or $1$ with $n_{i}=b_{i}^{\dag}b_{i}$.
Thus, we can use the bosonic version\cite{JPCM.18.9525,Phys.Rev.B.83.134507} of the CPT to calculate the dynamic spectrum of the model (\ref{eq2}).

In CPT, the original lattice is divided into identical clusters which constitute a superlattice. The lattice Hamiltonian is written as $H=H^{\prime}+V$,
where $H^{\prime}$ is the cluster Hamiltonian, obtained by severing the hopping terms between different clusters, which are now contained in $V$. The Green's function $\bm{G}$ of the original lattice is expressed (in matrix form) as\cite{Phys.Rev.Lett.84.522}
\begin{align}
\bm{G}(\tilde{\bm{k}},\omega)=\bm{G}^{\prime}(\omega)[1-\bm{V}(\tilde{\bm{k}})\bm{G}^{\prime}(\omega)]^{-1},
\end{align}
where $\bm{G}^{\prime}$ is the cluster Green's function calculated by the ED method\cite{Rev.Mod.Phys.66.763}, and $\tilde{\bm{k}}$ the wavevector in the BZ of the superlattice. Since the translation invariance of the original lattice is broken by the cluster decomposition, we use a periodization procedure to recover the translation invariance of Green's function\cite{Phys.Rev.Lett.84.522},
\begin{align}
G_{cpt}(\bm{k},\omega)=\frac{1}{N}\sum_{\mu,\nu}e^{-i\bm{k}\cdot(\bm{r}_{\mu}-\bm{r}_{\nu})}G_{\mu\nu}(\tilde{\bm{k}},\omega),
\end{align}
where $N$ is the number of the lattice sites in each cluster, $\mu$ and $\nu$ are the indices of the sites. Here, the wavevector $\bm{k}$ in the original BZ can be expressed as $\bm{k}=\tilde{\bm{k}}+\bm{K}$, where $\bm{K}$ is the reciprocal vector of the superlattice.

Since the intercluster coupling $V$ contains only the one-body terms, the NN and NNN interactions connecting different clusters can not be included into $V$. To treat these extended interactions, we perform a Hartree approximation\cite{PhysRevB.70.235107,PhysRevB.87.075123} and the intercluster interactions is then replaced by,
\begin{align}
J_{1}\sum_{\langle ij\rangle}(n_{i}\langle n_{j}\rangle+\langle n_{i}\rangle n_{j})
+J_{2}\sum_{\langle\langle ij\rangle\rangle}(n_{i}\langle n_{j}\rangle+\langle n_{i}\rangle n_{j}),
\end{align}
where $\langle n_{i}\rangle$ is the mean-field value of $n_{i}$ and is determined self-consistently.

In virtue of the mapping between $s=1/2$ spins and hard-core bosons, the dynamic spin susceptibility $\chi^{+-}(\bm{k},\omega)$ corresponds to the bosonic single-particle Green's function $G_{cpt}(\bm{k},\omega)$, so the spin structure factor is given by,
\begin{align}
\mathcal{S}^{+-}(\bm{k},\omega)=-\mathrm{Im}\chi^{+-}(\bm{k},\omega)=-\mathrm{Im}G_{cpt}(\bm{k},\omega).
\end{align}

To check the validity of the CPT method, we have applied it to the NN AF Heisenberg models on the chain and ladder, which have been well studied by analytical and numerical methods respectively\cite{PhysRev.128.2131,PTP.41.880,PhysRevB.88.094411}. We find that it can give the excitation spectra
not only for the magnon but also for the spinon and multimagnon excitations, which agree well with the analytical
results based on the Bethe Ansatz method or the numerical results based on the density-matrix-renormalization-group method\cite{SupMat}. In the following, the calculations for the square-lattice AF Heisenberg model are carried out by tiling the lattice with $4\times6$ clusters if not explicitly specified, and we have checked that the finite-size effects are very weak by comparing to the results with the $4\times4$ cluster tiling\cite{SupMat}.

\begin{figure}
  \centering
  \includegraphics[scale=0.9]{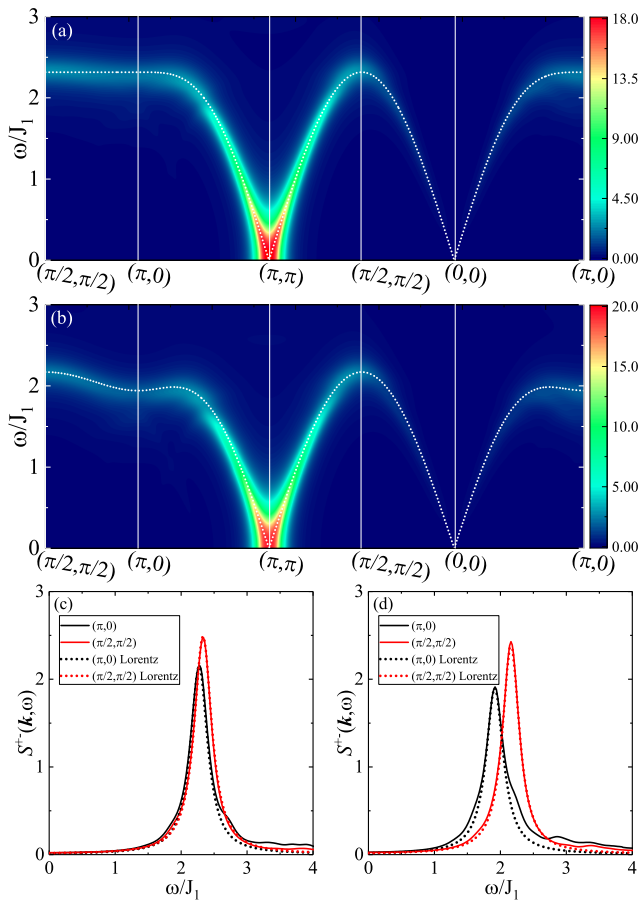}
  \caption{\label{spectra-AF}(a) and (b) are the dynamic structure factors for $J_{2}=0$ and $J_{2}=0.1J_{1}$, respectively. The white dotted lines are the linear spin-wave dispersions with $1/S$ correction. (c) and (d) are the dynamic structure factors at wave vectors $(\pi,0)$ and $(\pi/2,\pi/2)$ by the CPT calculation and Lorentz fitting for $J_{2}=0.0$ and $J_{2}=0.1J_{1}$, respectively.}
\end{figure}
 Let us first study the spectral properties in the N\'{e}el phase ($J_{2}\lesssim0.4J_{1}$). Fig.~\ref{spectra-AF}(a) shows the spin excitation spectra for the NN AF Heisenberg model ($J_{2}=0$) along several high symmetry lines.
Overall, the dispersion is in good agreement with that obtained by the spin-wave theory (SWT) with $1/S$ correction as denoted by the dotted lines\cite{PhysRevB.82.144407}. Especially, the CPT result can successfully produce the critical Goldstone mode at $(\pi,\pi)$.
However, the result with CPT shows a clear downward dispersion around $(\pi,0)$ so that a local minimum forms at that $\bm{k}$ point, while the SWT exhibits a nearly flat dispersion. This indicates that the spin excitations near $(\pi,0)$ deviate from the single-magnon modes which are the only excitations in the SWT. Figure \ref{spectra-AF}(c) shows the energy distributions of spectra at $(\pi,0)$ and $(\pi/2,\pi/2)$ and the fittings with the Lorentz functions denoted by the dotted lines. At $(\pi/2,\pi/2)$, the spectrum can be well fit by the Lorentz function, indicating that it comes essentially from the single-particle bosonic excitation, i.e., magnon. However, though the spectrum at $(\pi,0)$ is also dominated by the Lorentz-type magnon excitations, an additional long tail extended to high energies can be clearly observed, which exhibits as a continuum in the excitation spectra. By subtracting the weight of the Lorentz fittings from the total spectral weight, we find that the continuum accounts for $19\%$ of the total spectral weight at $(\pi,0)$ but only negligible $2\%$ at $(\pi/2,\pi/2)$. On the other hand, Fig.~\ref{spectra-AF}(c) shows that the excitation energy at $(\pi,0)$ is  lower than that at $(\pi/2,\pi/2)$, which can in fact also be discerned from the spectra shown in Fig.~\ref{spectra-AF}(a). Moreover, the spectral intensity at $(\pi,0)$ is smaller than that at $(\pi/2,\pi/2)$.

\begin{figure}
  \centering
  \includegraphics[scale=0.6]{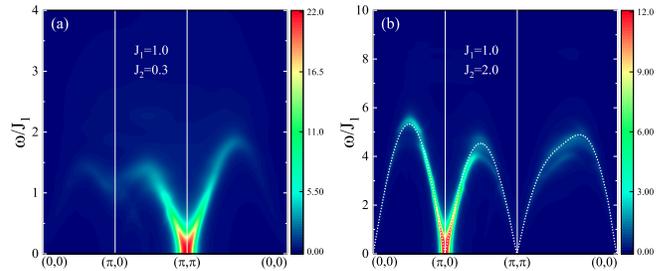}
  \caption{\label{neel-stripe}Dynamical structure factors for (a) $J_{2}=0.3J_{1}$ and (b) $J_{2}=2.0J_{1}$, respectively. The white dotted line in (b) is the linear spin-wave dispersion with $1/S$ correction.}
\end{figure}

These results presented in Fig.1 (a) and (c) with $J_2=0$ have already reproduced qualitatively the features of the anomalies near $(\pi,0)$ as observed by INS experiments on CFTD\cite{PhysRevLett.87.037202,pnas.0703293104,nphys3172}, namely the downward dispersion, an additional long tail spectral weights and the reduction of the spectral weight compared to that near $(\pi/2,\pi/2)$, indicating that our CPT calculation has captured more physics going beyond the usual SWT method.
However, the quantities of the anomalies are all obviously smaller than those observed in experiments\cite{PhysRevLett.87.037202,pnas.0703293104,nphys3172}. Considering that the NNN AF exchange can not be neglectable small in real materials, we turn on the NNN interaction $J_{2}$ and show the results for $J_{2}=0.1J_{1}$ in Fig.~\ref{spectra-AF}(b) and (d). Indeed, in comparison with the case of $J_{2}=0$, the spectra agree better with the experimental results\cite{PhysRevLett.87.037202,pnas.0703293104,nphys3172}. Specifically, now the continuum carries $28\%$ of the total spectral weights at $(\pi,0)$ but still only $5\%$ at $(\pi/2,\pi/2)$,
and the energy difference between $(\pi,0)$ and $(\pi/2,\pi/2)$ is increased to $0.23J_{1}$[see Fig.~\ref{spectra-AF}(d)]. These results indicate that a small $J_{2}$ is necessary to quantitatively reproduce the experimental observations, but the results for $J_{2}=0$ have already captured their main features. As we increase $J_{2}$ further, the quantities of the anomalies at $(\pi,0)$ is further enhanced as shown in Fig.~\ref{neel-stripe}(a) for $J_{2}=0.3J_{1}$.

We then turn to the spectra in another magnetically ordered stripe phase ($J_{2}\gtrsim0.6J_{1}$). The typical result for $J_{2}=2.0J_{1}$ is shown in Fig.~\ref{neel-stripe}(b). We find that the Goldstone mode shifts to $(\pi,0)$, which is consistent with the stripe order, and the low-energy dispersion is consistent with the SWT results denoted by the dotted lines\cite{PhysRevB.82.144407}. However, similar to the N\'{e}el phase for small $J_{2}$, there are also distinct continuums at high energies, and their positions are now near $(\pi/2,\pi/2)$ and $(\pi/2,\pi)$. Thus, the high-energy continuum is a common feature of the magnetically ordered states of the $J_{1}$-$J_{2}$ model, which suggests some kind of physics beyond the usual picture of magnon.

\begin{figure}
  \centering
  \includegraphics[scale=0.58]{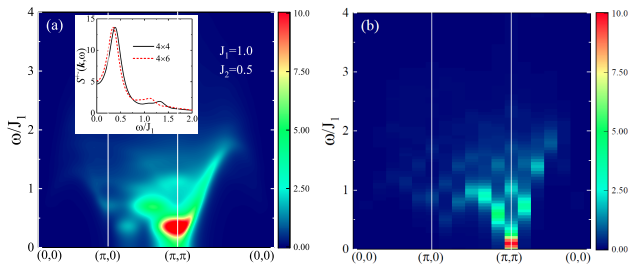}
  \caption{\label{spectra-qsl}(a) Dynamical structure factor for $J_{2}=0.5J_{1}$ calculated by CPT. The inset shows the spectra at $(\pi,\pi)$ for $4\times4$ and $4\times6$ cluster tilings. (b) Dynamical structure factor for $J_{2}=0.5J_{1}$ calculated by VMC with a $Z_{2}$ RVB mean-field state (see main text) on a $12\times12$ lattice.}
\end{figure}

Before discussing the origin of the non-spin-wave continuums, we will now study the nature of the magnetically disordered phase in the intermediate region ($0.4J_{1}\lesssim J_{2}\lesssim0.6J_{1}$), and the typical result for $J_{2}=0.5J_{1}$ is presented in Fig.~\ref{spectra-qsl}(a). One can see that the Goldstone mode related to magnetic orders disappears, and the whole spectrum becomes a broad continuum. This suggests strongly the fractionalized excitations coming from a QSL ground state. Moreover, we note that the spectrum is gapless, as can be seen more clearly from the inset of Fig.~\ref{spectra-qsl}(a).

In order to identify this possible QSL state, we resort to the VMC method\cite{PhysRevB.81.214509,SupMat} to carry out an analysis, based on the general resonant-valence-bond (RVB) mean-field (MF) Hamiltonian for spinons:
\begin{equation}
H_{0}=\sum_{\bm{k}\sigma}\varepsilon_{\bm{k}}f^\dagger_{\bm{k}\sigma}f_{\bm{k}\sigma}+\sum_{\bm{k}}(\Delta_{\bm{k}}f^\dagger_{\bm{k}\uparrow} f^\dagger_{-\bm{k}\downarrow}+h.c.).
\label{mf-spinon}
\end{equation}
The physical ground states for the spin model are represented by Gutzwiller-projected MF wave functions, and the $s=1$ excited states by the Gutzwiller-projected two-spinon wave functions.
As with the previous study\cite{PhysRevB.88.060402}, we find that a state with $\epsilon_{\bm{k}}=t(\cos{k_x}+\cos{k_y})$ and $\Delta_{\bm{k}}=\Delta_{1}(\cos{k_{x}}-\cos{k_{y}})+\Delta_{2}\sin{2k_{x}}\sin{2k_{y}}$ gives the best variational energy, and the optimal variational parameters are $\Delta_{1}=3.60t$ and $\Delta_{2}=2.20t$. More importantly, we find that the spin-excitation spectrum from the VMC calculation [Fig.~\ref{spectra-qsl}(b)] is consistent with the CPT results [Fig.~\ref{spectra-qsl}(a)] quite well. The only slight difference of the spectra between the two cases is that the maximum spectral weight at $(\pi,\pi)$ locates at a higher energy for the CPT result. However, from the inset of Fig.~\ref{spectra-qsl}(a), we can see that the spectral weight at $(\pi,\pi)$ will shift to low energies as the cluster size in CPT increases. These results indicate that the state obtain by the VMC is the most likely ground state for $J_{2}=0.5J_{1}$ and it is a gapless $Z_{2}$ QSL\cite{PhysRevB.88.060402} according to the theory of the projective-symmetry group\cite{PhysRevB.65.165113}.

\begin{figure}
  \centering
  \includegraphics[scale=0.6]{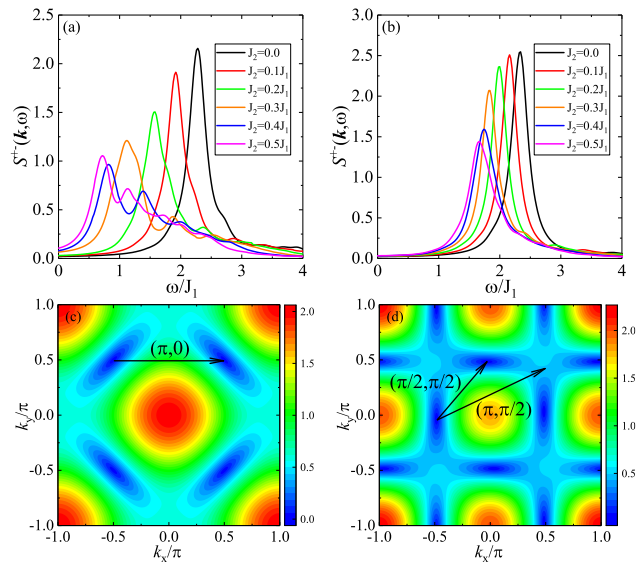}
  \caption{\label{evolution}(a) and (b) are the energy distributions of the dynamical structure factors for different $J_{2}$ at $\bm{k}=(\pi,0)$ and $(\pi/2,\pi/2)$, respectively. (c) and (d) are the contour plots of the dispersion of single-spinon excitations with the parameters obtained by VMC for $J_{2}=0.1J_{1}$ and $2.0J_{1}$, respectively.}
\end{figure}

 In Fig.~\ref{evolution}(a) and (b), we present the evolution of the spectra from $J_{2}=0$ to $J_{2}=0.5J_{1}$ at $(\pi,0)$ and $(\pi/2,\pi/2)$, respectively. First, we note that the high-energy continuums at $(\pi,0)$ are much stronger than those at $(\pi/2,\pi/2)$ for all of $J_{2}$. Most importantly, both Fig.~\ref{evolution}(a) and (b) show that the evolution of the spectra is continuous up to the $J_{2}=0.5J_{1}$ case with a QSL ground state. However, the evolution at $(\pi/2,\pi/2)$ exhibits mainly as the broadening of the Lorentz lineshape, while that
 at $(\pi,0)$ not only as the similar broadening but also as an enhancement of the continuum tail. As the elementary excitations in the QSL phase are fractionalized spinons carrying spin $s=1/2$, we ascribe the anomalous high-energy continuum in the magnetically ordered phases as the deconfinement of the spinons. In this way, the $s=1$ magnons are confined pair of spinons.

To understand why the fractionalized excitations appear around $(\pi,0)$ instead of other momentum points, we perform the same VMC analysis based on the MF Hamiltonian (\ref{mf-spinon}). For $J_{2}=0.1J_{1}$, we find the state with $\varepsilon_{\bm{k}}=t(\cos{k_{x}}+\cos{k_{y}})$, $\Delta_{\bm{k}}=\Delta(\cos{k_{x}}-\cos{k_{y}})$ and $\Delta=0.30t$ gives the optimal variational energy. The contour plot of the dispersion for a single-spinon excitation is presented in Fig.~\ref{evolution}(c), which shows four minima at $(\pm\pi/2,\pm\pi/2)$.
Thus, the minimum energy for two-spinon excitations will have wave vectors $(\pi,0)$ or $(\pi,\pi)$. But, the two spinons are confined around $(\pi,\pi)$ due to the existence of the N\'{e}el order, so the fractionalized excitations are most likely to appear around $(\pi,0)$.

This explanation also applies to the case of the stripe phase. For $J_{2}=2.0J_{1}$, we find that $\varepsilon_{\bm{k}}=t_{1}(\cos{k_{x}}+\cos{k_{y}})+2t_{2}\cos{k_{x}}\cos{k_{y}}$ and $\Delta_{\bm{k}}=\Delta_{1}(\cos{k_{x}}-\cos{k_{y}})+2\Delta_{2}\sin{k_{x}}\sin{k_{y}}$ with $t_{1}=-0.10t_{2}$, $\Delta_{1}=0.06t_{2}$ and $\Delta_{2}=-0.26t_{2}$ gives the optimal variational state. As shown in Fig.~\ref{evolution}(d), the two-spinon excitations at $(\pi/2,\pi/2)$ have the lowest excitation energy and those at $(\pi,\pi/2)$ have a similar lower excitation energy, so the deconfinement of spinons emerges at these momenta. While a pair of spinons are confined to form a magnon around $(\pi,0)$ and $(\pi,\pi)$ due to the existence of the stripe order, though the two-spinon excitation also carries a minimum energy at these momenta, which is similar to the N\'{e}el phase as described above.

In summary, we extend the cluster perturbation theory to the $s=1/2$ spin model, and study the spin excitation spectra for the AF $J_{1}$-$J_{2}$ Heisenberg model on the square lattice. We show clearly the confinement and deconfinement of the spin-1/2 spinons with a variation of $J_{2}$. In the magnetically ordered phases, the spin excitations are partially fractionalized at hight energies, though the low-energy excitations are well defined magnons. In the magnetically disordered phase, the whole spectrum becomes a broad continuum suggesting a complete
fractionalization of spin excitations, whose ground state is ascribed to be a $Z_{2}$ quantum spin liquid.
Our investigation also highlights that the developed approach is applicable to a variety of long-range ordered and disordered (spin liquid) phases.

\begin{acknowledgments}
This work was supported by the National Natural Science Foundation of China (11674158, 11404163 and 11774152), National Key Projects for Research and Development
of China (Grant No. 2016YFA0300401) and Natural Science Foundation of Jiangsu Province (BK20140589). W. W. was also supported by the program B for Outstanding
PhD candidate of Nanjing University.
\par S.-L. Y. and W. W. contributed equally to this work.
\end{acknowledgments}

\bibliography{j1-j2}

\appendix

\begin{center}
\textbf{\large
Supplementary Material for: ``Spectral properties of the square-lattice antiferromagnetic $J_{1}$-$J_{2}$ Heisenberg model: confinement and deconfinement of spinons"}
\end{center}

Here, we show the cluster-perturbation-theory (CPT) results of the antiferromagnetic (AF) Heisenberg model on the one-dimension chain and ladder, the CPT results of the $J_{1}$-$J_{2}$ model with $4\times4$ cluster tiling, and the details about the variational-Monte-Carlo (VMC) methods used in the main text.

\section{CPT results on the one-dimension chain and ladder}

\begin{figure}
  \centering
  \includegraphics[scale=0.57]{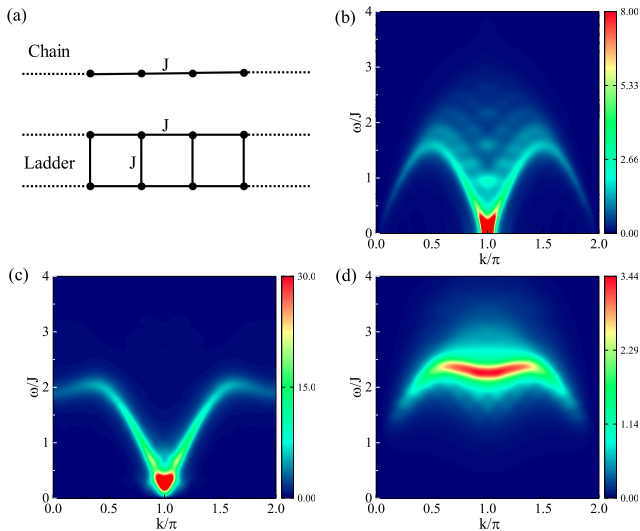}
  \caption{\label{chain-ladder} (a) Schematic diagrams of the one-dimensional and ladder Heisenberg models. (b) Dynamical structure factor $\mathcal{S}^{+-}(\bm{k},\omega)$ for the one-dimensional AF Heisenberg model. (c) Antisymmetric dynamical structure factor $\mathcal{S}^{+-}_{\pi}(\bm{k},\omega)$ for the AF Heisenberg model on the ladder. (d) Symmetric dynamical structure factor $\mathcal{S}^{+-}_{0}(\bm{k},\omega)$ for the AF Heisenberg model on the ladder.}
\end{figure}
Figure~\ref{chain-ladder}(a) illustrates the schematic diagrams of the one-dimensional and ladder Heisenberg models. In the cluster-perturbation-theory (CPT) calculations, we use $1\times12$ and $2\times12$ clusters for the chain and ladder models, respectively. In the one-dimensional chain, the elementary excitations are the fractionalized $s=1/2$ spinons. The dynamical structure factor $\mathcal{S}^{+-}(\bm{k},\omega)$ for the one-dimensional AF Heisenberg model is shown in Fig.~\ref{chain-ladder}(b), in which the obvious two-spinon continuum spectrum can be seen. This is the characteristic of the fractionalization of the spin excitation, and it is consistent with the analytical results based on the Bethe Ansatz method\cite{PhysRev.128.2131,PTP.41.880}. In the ladder system, the spinon excitations of the individual chains are confined by the interchain couplings. Thus, the gapless two-spinon continuum in the one-dimensional chain is replaced by the gapped $s=1$ two-spinon bound states, which are referred to as ``magnons". For the ladder system, the momentums along the rung direction have only two values, i.e., $k_{y}=0$ and $k_{y}=\pi$, so we can define two types of dynamical structure factors $\mathcal{S}^{+-}_{0}(\bm{k},\omega)$ and $\mathcal{S}^{+-}_{\pi}(\bm{k},\omega)$, which are symmetric ($k_{y}=0$) and antisymmetric ($k_{y}=\pi$) for exchanging two legs of the ladder, respectively. Moveover, odd and even number of magnon excitations contribute to
the asymmetric and symmetric channel\cite{PhysRevB.88.094411}, respectively. The results for $\mathcal{S}^{+-}_{\pi}(k,\omega)$ and $\mathcal{S}^{+-}_{0}(\bm{k},\omega)$ are shown in Fig.~\ref{chain-ladder}(c) and (d). The excitation spectrum shown in Fig.~\ref{chain-ladder}(c) is made up of the low-energy gapped one-magnon excitations, and it exhibits a sharp structure. On the other hand, the spectrum in the symmetric channel shown in Fig.~\ref{chain-ladder}(d) is a continuum, which is resulted from the two-magnon excitations. These results of the spin ladder agree well with those obtained by the density-matrix-renormalization-group method\cite{PhysRevB.88.094411}.

\section{CPT results for the $J_{1}$-$J_{2}$ model with $4\times4$ cluster tiling}

\begin{figure}
  \centering
  \includegraphics[scale=0.57]{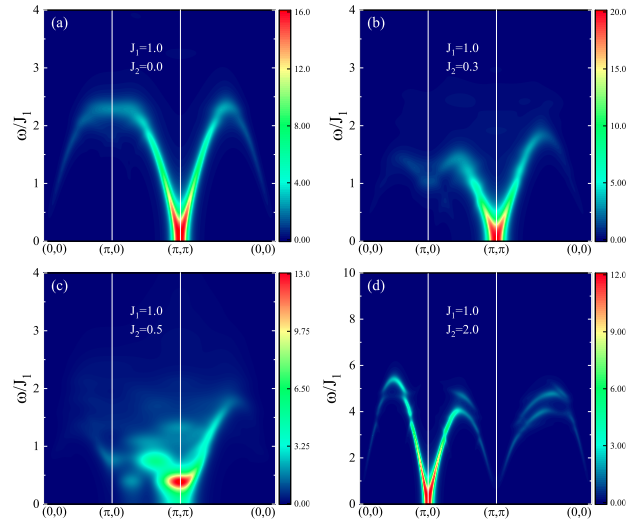}
  \caption{\label{cpt-4-by-4} Dynamical structure factors for $J_{1}$-$J_{2}$ model calculated by CPT with the $4\times4$ cluster tiling}
\end{figure}
To check the finite-size effects of the clusters in the CPT, we here present the results calculated from the $4\times4$ cluster tiling (see Fig.~\ref{cpt-4-by-4}), in addition to the $4\times6$ cluster tiling used in the main text. We find that the spin-excitation spectra in Fig.~\ref{cpt-4-by-4} are in good agreement with those in the main text for the $4\times6$ cluster tiling, which implies that the finite-size effects are very weak for the $4\times6$ cluster tiling.

\section{Details of the VMC method}

We briefly introduce the VMC method for calculating the ground state and the dynamical excitation spectra of an $s=1/2$ spin model.

In order to describe the fractionalization of spin excitations, a natural way is to use a fermion representation of the spin operator:
\begin{align}
S^{\mu}_{i}=\frac{1}{2}c^{\dag}_{i\sigma}\tau^{\mu}_{\sigma\sigma^{\prime}}c_{i\sigma^{\prime}},
\end{align}
where $c^{\dag}_{i\sigma}$ ($c_{i\sigma}$) creates (annihilates) a spinon at site $i$ with spin $\sigma$ and $\tau^{\mu}_{\alpha\beta}$ ($\mu=1,2,3$) are the Pauli matrices. With this fermion representation, the $J_{1}-J_{2}$ model then becomes
\begin{align}
H&=J_{1}\sum_{\langle ij\rangle}[\frac{1}{4}(n_{i\uparrow}-n_{i\downarrow})(n_{j\uparrow}-n_{j\downarrow}) \nonumber \\
&+\frac{1}{2}(c^{\dag}_{i\uparrow}c_{i\downarrow}c^{\dag}_{j\downarrow}c_{j\uparrow}+c^{\dag}_{i\downarrow}c_{i\uparrow}c^{\dag}_{j\uparrow}c_{j\downarrow})] \nonumber \\
&+J_{2}\sum_{\langle\langle ij\rangle\rangle}[\frac{1}{4}(n_{i\uparrow}-n_{i\downarrow})(n_{j\uparrow}-n_{j\downarrow}) \nonumber \\
&+\frac{1}{2}(c^{\dag}_{i\uparrow}c_{i\downarrow}c^{\dag}_{j\downarrow}c_{j\uparrow}+c^{\dag}_{i\downarrow}c_{i\uparrow}c^{\dag}_{j\uparrow}c_{j\downarrow})].
\end{align}
In this representation, we have enlarge the Hilbert space, so a constraint on each site must be considered:
\begin{align}
\sum_{\sigma}c^{\dag}_{i\sigma}c_{i\sigma}=1.
\label{single-constraint}
\end{align}

By using the VMC method, we can study the various quantum-spin-liquid (QSL) states and magnetically ordered states. Our starting point is the mean-field (MF) Hamiltonian:
\begin{equation}
H_{0}=\sum_{\bm{k}\sigma}\varepsilon_{\bm{k}}c^\dagger_{\bm{k}\sigma}c_{\bm{k}\sigma}+\sum_{\bm{k}}(\Delta_{\bm{k}}c^\dagger_{\bm{k}\uparrow} c^\dagger_{-\bm{k}\downarrow}+h.c.).
\label{mf-spin-liquid}
\end{equation}
The physical ground states for the spin model are represented by
\begin{equation}
|\Psi\rangle=P_{G}P_{N}|\Psi_{0}\rangle,
\end{equation}
in which $|\Psi_{0}\rangle$ is the ground state of the MF Hamiltonian $H_{0}$, $P_{G}=\prod_{i}(1-n_{i\uparrow}n_{i\downarrow})$ is the Gutzwiller projection operator to enforce the constraint (\ref{single-constraint}) and the operator $P_{N}$ projects the state into the subspace with $N$ spinons. The optimal parameters in $\varepsilon_{\bm{k}}$ and $\Delta_{\bm{k}}$ in Eq.~(\ref{mf-spin-liquid}) is determined by minimizing the energy expectation value
\begin{equation}
E[\Psi]=\frac{\langle\Psi|H|\Psi\rangle}{\langle\Psi|\Psi\rangle},
\end{equation}
which can be evaluated by the Monte Carlo method.

The dynamical structure factor can be written as
\begin{equation}
S^{+-}(\bm{q},\omega)=\frac{1}{\langle\Psi|\Psi\rangle}\sum_{n}|\langle n|S^{+}_{\bm{q}}|\Psi\rangle|^2\delta[\omega-(E_n-E_g)].
\end{equation}
Here, $|n\rangle$ is the exited state with energy $E_n$ and $E_g$ is the energy of ground state. Due to the commutability between the spin operator $S^{+}_{\bm{q}}$ and the operators $P_{G}$ and $P_{N}$, we have
\begin{equation}
S^{+}_{\bm{q}}|\Psi\rangle=P_{G}P_{N}S^{+}_{\bm{q}}|\Psi_{0}\rangle=\sum_{k}|\bm{k},\bm{q}\rangle,
\end{equation}
where
\begin{equation}
|\bm{k},\bm{q}\rangle=P_{G}P_{N}c^{\dag}_{\bm{k}+\bm{q}\uparrow}c_{\bm{k}\downarrow}|\Psi_{0}\rangle.
\end{equation}
Then, we can project the physical Hamiltonian $H$ into the subspace spanned by the states $|\bm{k},\bm{q}\rangle$ and express the excited states as $|n\rangle=\sum_{\bm{k}}\phi^{n}_{\bm{k}\bm{q}}|\bm{k},\bm{q}\rangle$. In order to determine the coefficients $\phi^{n}_{\bm{k}\bm{q}}$
and energies $E_{n}$, we diagonalize the projected Hamiltonian by solving a generalized eigenvalue problem
\begin{equation}
\sum_{\bm{k}^\prime}H_{\bm{k}\bm{k}^{\prime}}^{\bm{q}}\phi^{n}_{\bm{k}^{\prime}\bm{q}}
=E_{n}\sum_{\bm{k}^\prime}\mathcal{O}^{\bm{q}}_{\bm{k}\bm{k}^{\prime}}\phi^n_{\bm{k}^{\prime}\bm{q}},
\end{equation}
with $H_{\bm{k}\bm{k}^{\prime}}^{\bm{q}}=\langle \bm{k},\bm{q}|H|\bm{k}^{\prime},\bm{q}\rangle$ and $\mathcal{O}^{\bm{q}}_{\bm{k}\bm{k}^{\prime}}=\langle\bm{k},\bm{q}|\bm{k}^{\prime},\bm{q}\rangle$. These matrices are calculated by the Monte Carlo reweighing technique\cite{PhysRevB.81.214509} in which the sampling probability becomes ${|\langle \bm{k},\bm{q}|\alpha\rangle|^2}/\sum_{\bm{k}}\mathcal{O}^{\bm{q}}_{\bm{k}\bm{k}}$ with $|\alpha\rangle$ being the real space spin configuration.

\end{document}